\documentclass[10pt]{article}
\pdfoutput=1
\usepackage{amssymb,amsmath,mathrsfs,enumerate}
\usepackage{graphicx,rotate,multicol}
\usepackage{wrapfig}
\usepackage[colorlinks=true,
            linkcolor=red,
            urlcolor=blue,
            citecolor=blue]{hyperref}
\usepackage{color}
\usepackage{soul}
\usepackage{cite}
\usepackage{bbold}
\usepackage{verbatim}
\long\def\rpl#1!!#2!!{\textcolor{red}{#1} \textcolor{blue}{#2}}

\DeclareMathOperator{\Tr}{Tr}

\def \order(#1){{\cal O} \left(#1 \right)}

\evensidemargin=\oddsidemargin
\def\Eqn#1{Eq.\ (\ref{#1})}

\allowdisplaybreaks

\begin{document}
\begin{flushright}
    LU TP 18-35
\end{flushright}

\begin{center}
	{\Large \bf Cornering variants of Georgi-Machacek model \\ using Higgs precision data } \\
	\vspace*{1cm} {\sf Dipankar
          Das$^{a,b,}$\footnote{dipankar.das@thep.lu.se},~Ipsita
          Saha$^{c,}$\footnote{ipsita.saha@roma1.infn.it}} \\
	\vspace{10pt} {\small \em 
          $^a$Department of Physics, University of Calcutta, 92
          Acharya Prafulla Chandra Road, Kolkata 700009, India \\
	      $^b$Department of Astronomy and Theoretical Physics, Lund University,
	       221 00 Lund, Sweden \\
		$^c$Istituto Nazionale di Fisica Nucleare, Sezione di Roma,
		Piazzale Aldo Moro, 2, 00185, Roma, Italy }
	
	\normalsize
\end{center}

\begin{abstract}
We show that in the absence of trilinear terms in the scalar potential 
of Georgi-Machacek model, heavy charged
scalars do not necessarily decouple from the $h\to \gamma\gamma$ decay amplitude.
In such scenarios, measurement of the Higgs to diphoton signal strength can place
stringent constraints in the parameter space. Using the projected precisions at the
High Luminosity LHC (HL-LHC) and the ILC, we find that the upper bound on the triplet
vacuum expectation value can be as low as $10$~GeV.
We also found that when combined with the theoretical constraints from perturbative
unitarity and stability, such variants may be ruled out altogether.
\end{abstract}

\bigskip

%
The discovery of a Higgs-like scalar at the Large Hadron Collider~(LHC)
\cite{Aad:2012tfa,Chatrchyan:2012xdj}, 
has initiated a deeper investigation into its origin. It is now time
to settle whether this new particle is the Higgs boson as predicted by the Standard Model~(SM),
or it is a Higgs-like particle stemming from a more elaborate construction beyond the SM~(BSM).
In view of the latter intriguing possibility, recent years have seen a growing interest in BSM scenarios with an extended Higgs
sector, where the scalar boson observed at the LHC is only the first to appear in a series of many
others to follow. The Georgi-Machacek~(GM) model\cite{Georgi:1985nv,Chanowitz:1985ug,Gunion:1989ci}
constitute one of the simplest examples of
this category and therefore have received a lot of attention  in recent times.

The SM extended by a scalar triplet only, which goes by the name of Higgs triplet
model~(HTM) does not preserve the custodial symmetry at the tree-level. Consequently,
the vacuum expectation value~(VEV) of the triplet is restricted to be below $\order(10~{\rm GeV})$\cite{Kanemura:2012rs}
so that the value of the electroweak $\rho$-parameter remains consistent with the experimental
fits. In the GM model, two scalar triplets (one real and one complex) are added in such a
way that the custodial symmetry is preserved at the tree-level in the limit when the
two triplets assume a common VEV, $v_t$, after the spontaneous symmetry breaking~(SSB). Therefore, unlike in the HTM, the constraint on $v_t$ arising
from the electroweak $\rho$-parameter is lifted in the case of GM model.

However, since the charged scalars couple to the SM fermions with strengths
proportional to $v_t$, upper bound on $v_t$ can still be placed using the flavor data
when the charged scalars are sufficiently light. But such an upper
bound gets diluted very quickly with increasing charged scalar masses to the extent that
for charged scalar masses of $\order(1 ~{\rm TeV})$ one can allow $v_t$ to be as
large as $60$~GeV\cite{Hartling:2014aga,Biswas:2018jun}.
Our motivation in this paper is to put constraints on the triplet VEV, $v_t$, even
when the nonstandard scalars are super heavy. We will employ the Higgs data to
achieve this purpose. As a first step, we shall reanalyze the scalar potential
of the GM model and identify the lightest CP-even scalar~($h$) in the spectrum
as the $125$~GeV resonance observed at the LHC. Then we compute the $VVh$ ($V=W,Z$)
and $\bar{f}fh$ ($f$ denotes a generic fermion) in terms of the model parameters
and compare them with the corresponding SM expectations taking into account
current and future experimental accuracies. Admittedly, such studies have been
performed earlier in the literature\cite{Logan:2010en,Kanemura:2013mc,Belanger:2013xza,Hartling:2014zca,Chiang:2014bia,Kanemura:2014bqa,Chiang:2015kka,Logan:2015xpa,Chiang:2015rva, Chiang:2015amq,Chang:2017niy,Li:2017daq} but our current analysis differs
in a crucial way in the sense that we highlight how the inclusion of $h\to\gamma\gamma$
signal strength measurement can significantly tighten the constraints in the parameter
space. One key observation of our paper, which runs contrary to the common perception\cite{Hartling:2014zca}
is that the charged scalar contributions to the $h\to\gamma\gamma$ decay amplitude
do not necessarily decouple in certain variants of the GM models. Consequently, for
such variants, the bounds on the model parameters, especially $v_t$, extracted using
the data for Higgs signal strengths will not get washed away even when the nonstandard
masses are in the multi-TeV range. To our knowledge, this subtle but interesting feature
has not been emphasized earlier in the context of GM models.

%
We start by revisiting the scalar potential of the GM model. As stated earlier,
GM model~\cite{Georgi:1985nv,Chanowitz:1985ug,Gunion:1989ci} extends the scalar sector of the SM by adding one real $SU(2)_L$ triplet $\xi$ 
and one complex $SU(2)_L$ triplet $\chi$ with hypercharges  $Y=0$  and $Y=2$ respectively.  The scalar fields of such a model can be concisely
represented in the forms of a bi-doublet $\Phi$ and a bi-triplet $X$ as
\begin{eqnarray}
\label{e:fields}
\Phi = \left(\begin{array}{cc}
\phi^{0*} & \phi^+ \\
-\phi^- & \phi^0 \\ 
\end{array}\right) \,, \qquad
X = \left(\begin{array}{ccc}
\chi^{0*} & \xi^+ & \chi^{++} \\
-\chi^- & \xi^0 & \chi^+ \\
\chi^{--} & -\xi^- & \chi^0 \\
\end{array}\right) \,.
\end{eqnarray}

In order to preserve the custodial symmetry, both the real and complex triplet should acquire the same VEV.
 Accordingly, the VEVs of the scalar multiplets are defined by 
 $\left<\Phi\right> = (v_d/\sqrt{2}) \mathbb{1}_{2\times2}$
and $\left<X\right> = v_t \mathbb{1}_{3\times3}$. 
Comparing the $W$ and $Z$ boson masses with their corresponding SM
expressions, we obtain the formula for  the electroweak~(EW) VEV as
\begin{eqnarray}
v= \sqrt{v_d^2 + 8 v_t^2} = 246~{\rm GeV} \,.
\label{e:custodial}
 \end{eqnarray}
Following the convention of Refs.~\cite{Aoki:2007ah,Chiang:2012cn,Hartling:2014zca,Degrande:2017naf}, we now write
down the most general scalar potential for the GM model as follows:
\begin{eqnarray}
V(\Phi,X) &=& \frac{\mu_2^2}{2} \Tr(\Phi^\dagger \Phi) + \frac{\mu_3^2}{2} \Tr(X^\dagger X) + \lambda_1[\Tr(\Phi^\dagger \Phi)]^2 +
\lambda_2 \Tr(\Phi^\dagger \Phi)\Tr(X^\dagger X) \nonumber \\ 
&& + \lambda_3 \Tr(X^\dagger X X^\dagger X) + \lambda_4 [\Tr(X^\dagger X)]^2 - \lambda_5 \Tr(\Phi^\dagger \tau_a \Phi \tau_b)\Tr(X^\dagger t_a X t_b) \nonumber \\
&& - M_1\Tr(\Phi^\dagger \tau_a \Phi \tau_b)\left( U X U^\dagger \right)_{ab} - M_2\Tr(X^\dagger t_a X t_b)\left( U X U^\dagger \right)_{ab} \,,
\label{e:potential}
\end{eqnarray}
where, $\tau_a \equiv \sigma_a/2$, ($a=1,2,3$) with $\sigma_a$s  being 
the Pauli matrices and  $t_a$s correspond to the generators of triplet representation of
 $SU(2)$ and are expressed as
\begin{eqnarray}
t_1 = \frac{1}{\sqrt{2}}\left(\begin{array}{ccc}
0 & 1 & 0 \\
1 & 0 & 1 \\
0 & 1 & 0 \\
\end{array}\right)\,, \qquad
t_2 = \frac{1}{\sqrt{2}}\left(\begin{array}{ccc}
0 & -i & 0 \\
i & 0 & -i \\
0 & i & 0 \\
\end{array}\right)\,, \qquad
t_3 =\left(\begin{array}{ccc}
1 & 0 & 0 \\
0 & 0 & 0 \\
0 & 0 & -1 \\
\end{array}\right)\,.
\label{su2_tripgen}
\end{eqnarray}
The matrix $U$ appearing in the cubic terms of \Eqn{e:potential} is given by,
\begin{eqnarray}
U = \frac{1}{\sqrt{2}} \left(\begin{array}{ccc}
-1 & 0 & 1 \\
-i & 0 & -i \\
0 & \sqrt{2} & 0 \\
\end{array}\right) \,.
\label{matU}
\end{eqnarray}
In the limit when both the triplets receive the same VEV, we will have two
independent minimization conditions, using which we can trade the bilinear
coefficients, $\mu_2^2$ and $\mu_3^2$ in favor of $v_d$ and $v_t$ as
follows:
\begin{subequations}
	\begin{eqnarray}
	\mu_2^2 &=& \frac{3}{2} M_1 v_t - 4 \lambda_1 v_d^2 - 3\left(2\lambda_2 -\lambda_5\right) v_t^2  \,, \\
	\mu_3^2 &=& \frac{M_1 v_d^2}{4v_t} + 6 M_2 v_t-\left(2\lambda_2 -\lambda_5\right) v_d^2 - 4 \left(
	\lambda_3+ 3\lambda_4 \right)v_t^2  \,.
	\end{eqnarray}
	\label{e:bilinears}
\end{subequations}
%
%
After SSB, the neutral component of the fields are expanded around the minima as
%
\begin{eqnarray}
\phi^0 = \frac{v_d}{\sqrt{2}} + \frac{h_d + i z_d}{\sqrt{2}}\,, \qquad  
\xi^0 = v_t + h_\xi \,, \qquad 
\chi^0 = v_t + \frac{h_\chi + i z_\chi}{\sqrt{2}} \,.
\label{e:EWSB}
\end{eqnarray}
%
The mass terms in the scalar potential, then, can be diagonalized to obtain a
custodial quintuplet $(H_5^{++},H_5^+,H_5^0, H_5^-, H_5^{--})$ with
common mass $m_5$, a custodial triplet $(H_3^+,H_0, H_3^-)$ with common
mass $m_3$ and two custodial singlets $h$ and $H$ with masses $m_h$
and $m_H$ respectively.\footnote{For details of the diagonalization
procedure we refer the reader to Refs.~\cite{Chiang:2012cn,Hartling:2014zca}.} In particular, to diagonalize
the $CP$-even scalar sector, we need to introduce an angle $\alpha$ which
is defined as follows
\begin{eqnarray}
\label{e:alpha}
\begin{pmatrix} H_5^0 \\ H  \\ h \end{pmatrix} =
\begin{pmatrix}
1 & 0 & 0 \\ 0 & \cos\alpha & -\sin\alpha \\ 0 & \sin\alpha & \cos\alpha
\end{pmatrix}
\begin{pmatrix}
\sqrt{\frac{2}{3}} & -\sqrt{\frac{1}{3}} & 0  \\
\sqrt{\frac{1}{3}} & \sqrt{\frac{2}{3}} & 0 \\
0 & 0 & 1
\end{pmatrix}
\begin{pmatrix} h_\xi \\ h_\chi  \\ h_d \end{pmatrix} \,.
\end{eqnarray}
In what follows, we assume $h$ to be the lightest among the $CP$-even scalars,
which has been discovered at the LHC with mass $m_h \approx 125$~GeV.

At this point it is instructive to count the number of independent parameters in
the potential of \Eqn{e:potential}. There are a total of nine parameters consisting
of two bilinears ($\mu_2^2$ and $\mu_3^2$), five quartic couplings ($\lambda_i$,
$i=1,\dots, 5$) and two cubic couplings ($M_1$ and $M_2$). Among these, 
$\mu_2^2$ and $\mu_3^2$ have already been traded in favor of $v_d$ and $v_t$
as in \Eqn{e:bilinears}. The five quartic couplings can now be exchanged in favor
of four physical masses, $m_5$, $m_3$, $m_H$ and $m_h$ and the mixing angle,
$\alpha$. The relevant formulas connecting the $\lambda_i$-s with the physical
masses and mixings are given below:
\begin{subequations}
	\begin{eqnarray}
	\lambda_1 &=& \frac{1}{8 v^2 \cos^2 \beta}\left(m_h^2 \cos^2 \alpha +m_H^2 \sin^2 \alpha\right) \,, \label{lambda1} \\
		\lambda_2 &=&\frac{1}{12 v^2 \cos \beta \sin \beta}\left( \sqrt{6}\left(m_h^2  - m_H^2\right) \sin 2 \alpha - 3 \sqrt{2}v \cos\beta M_1 + 12 m_3^2 \sin\beta \cos\beta\right) \,, \label{lambda2}\\	
\lambda_3 &=& \frac{1}{v^2 \sin^2 \beta}\left(\sqrt{2}v \cos\beta \cot \beta M_1 - 3 \sqrt{2} v \sin\beta M_2 - 3 m_3^2 \cos^2\beta +m_5^2 \right)  \,, \label{lambda3} \\
\lambda_4 &=& \frac{1}{6 v^2 \sin^2 \beta}\Big( 2 m_H^2 \cos^2 \alpha + 2 m_h^2 \sin^2\alpha - 2 m_5^2 
+6 \cos^2\beta m_3^2  \nonumber \\
&&   - 3\sqrt{2} v \cos\beta \cot \beta M_1 + 9 \sqrt{2} \sin \beta v M_2 \Big) \,, \label{lambda4} \\
	\lambda_5 &=& -\frac{ \sqrt{2} M_1}{ v \sin \beta} + \frac{2 m_3^2}{v^2} \,. \label{lambda5}
		\end{eqnarray}
		\label{e:masstolam}
\end{subequations}
where, we have defined
\begin{eqnarray}
\tan\beta = \frac{2\sqrt{2} v_t}{v_d} \,.
\end{eqnarray}
Note that \Eqn{e:masstolam} will be extremely useful in translating the scalar
couplings as well as our final results in terms of the physical parameters which
are directly accessible at the experiments.

\begin{figure}[htbp!]
    \centering
    \includegraphics[scale=0.3]{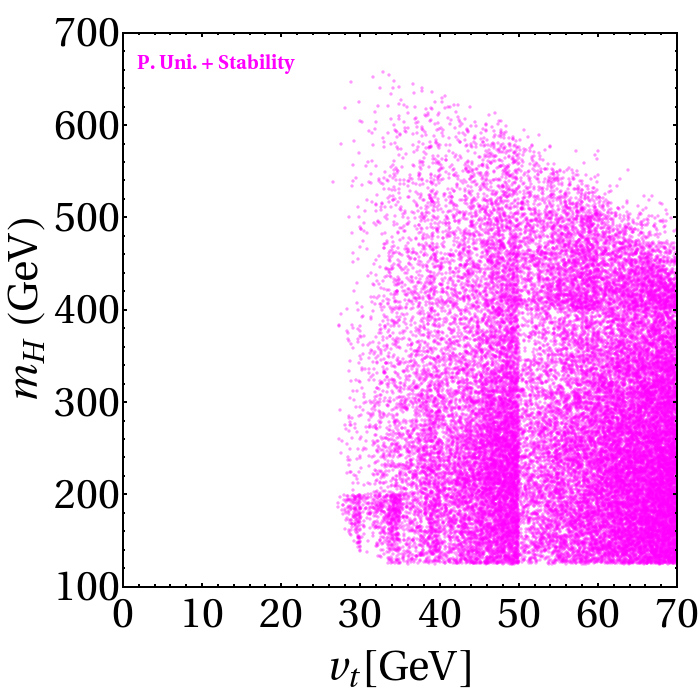}
    \caption{\em Allowed region in the $v_t$-$m_H$ plane from perturbative
        unitarity and stability in the limit $M_1,M_2=0$. Note that a lower
        limit on $v_t$ emerges from these theoretical constraints.}
    \label{f:uni}
\end{figure}

Before moving on to the main part of the paper, let us first examine
the theoretical bounds coming from perturbative unitarity and stability
of the scalar potential. The relevant constraints have been summarized
in Refs.~\cite{Aoki:2007ah,Hartling:2014zca}. In Fig.~\ref{f:uni} we display our main results in the
limit $M_1,M_2=0$. It is quite intriguing to note that there arises
a lower bound on $v_t$ ($\gtrsim 27$~GeV) from unitarity and stability.
It is not very difficult to find a qualitative justification for such
a lower bound as we will explain below.

Among the many unitarity constraints, we draw the reader's attention to
the following ones\cite{Aoki:2007ah,Hartling:2014zca} in particular:
\begin{eqnarray}
	|16\lambda_3+ 8\lambda_4| &\le& 8\pi \,, \label{e:uni1} \\
	|4\lambda_3+ 8\lambda_4| &\le& 8\pi  \label{e:uni2} \,.
\end{eqnarray}
After using \Eqn{e:masstolam} to substitute for the $\lambda_i$-s
and putting $M_1,M_2=0$, the above conditions read
\begin{eqnarray}
\left|\frac{5}{3} m_5^2 -5m_3^2\cos^2\beta +\frac{1}{3} m_0^2 \right| &\le& \pi v^2\sin^2\beta \,, \label{e:m1} \\
\left|\frac{1}{3} m_5^2 -m_3^2\cos^2\beta +\frac{2}{3} m_0^2 \right| &\le& 2\pi v^2\sin^2\beta \label{e:m2} \,.
\end{eqnarray}
where we have used the following abbreviation:
\begin{eqnarray}
	m_0^2 = m_h^2\sin^2\alpha +m_H^2\cos^2\alpha \,.
\end{eqnarray}
Clearly, for very small $v_t$, {\it i.e.}, $\sin\beta \to 0$, the following correlations
will be imposed:
\begin{eqnarray}
\frac{5}{3} m_5^2 -5m_3^2 +\frac{1}{3} m_0^2 &\approx& 0 \,, \label{e:cor1} \\
\frac{1}{3} m_5^2 -m_3^2 +\frac{2}{3} m_0^2 &\approx& 0 \label{e:cor2} \,.
\end{eqnarray}
Moreover, the following unitarity conditions will also be relevant\cite{Aoki:2007ah,Hartling:2014zca}:
\begin{eqnarray}
	|x_2^\pm| \equiv \left|4\lambda_1-2\lambda_3+4\lambda_4 \pm 
	\sqrt{(4\lambda_1+2\lambda_3-4\lambda_4)^2 +4\lambda_5^2} \right| \le 8\pi \,.
\end{eqnarray}
Using the triangle inequality we may write,
\begin{subequations}
	\begin{eqnarray}
		&& |x_2^+ +x_2^-| \le 16\pi \\
\Rightarrow	&&	|4\lambda_1-2\lambda_3+4\lambda_4| \le 8\pi \,.
	\end{eqnarray}
\end{subequations}
In the above inequality, the terms involving $\lambda_3$ and $\lambda_4$ will give
rise to the dominant contributions on the left hand side for small $v_t$. Again,
using \Eqn{e:masstolam} one can check that the limit $\sin\beta\to 0$ will entail
the following correlation:
\begin{eqnarray}
\frac{5}{3} m_5^2 -5 m_3^2 -\frac{2}{3} m_0^2 &\approx& 0 \label{e:cor3} \,.
\end{eqnarray}
It is easy to verify that no nontrivial solution set $\{m_5^2,m_3^2,m_0^2\}$ exists
such that Eqs.~(\ref{e:cor1}), (\ref{e:cor2}) and (\ref{e:cor3}) are satisfied
simultaneously. This explains why there should be a lower limit on $v_t$ (or
equivalently, on $\tan\beta$) so that the correlations of Eqs.~(\ref{e:cor1}), 
(\ref{e:cor2}) and (\ref{e:cor3}) do not become too strong.

\begin{figure}[htbp!]
    \begin{minipage}{0.46\textwidth}
        \centerline{\includegraphics[scale=0.3]{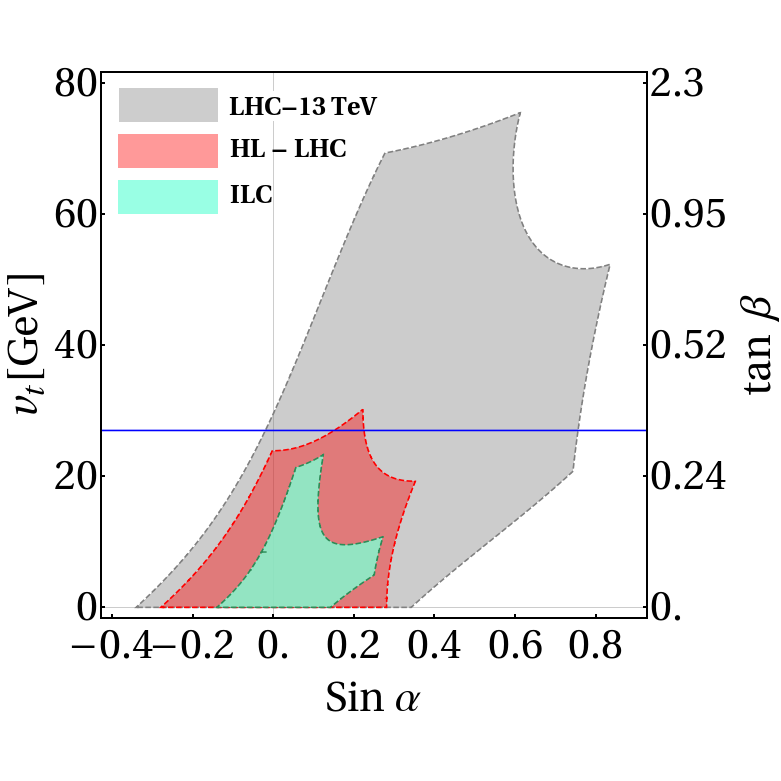}}
        \caption{\em Combined allowed regions at 95\% C.L. in the $\sin\alpha$-$v_t$ plane using the current\cite{CMS:2018lkl}
            (shaded in gray) and projected\cite{Fujii:2015jha} (shaded in red and green) accuracies in the measurements of $\kappa_V$ and $\kappa_f$.
            While extracting bounds using future precisions, the central values of all the $\kappa_x$
            parameters have been assumed to be $1$, {\it i.e.}, consistent with the SM.
            The horizontal line corresponds to the lower limit on $v_t$ arising from
            the theoretical constraints.}
        \label{f:kvkf}
    \end{minipage}
    \hfill
    \begin{minipage}{0.46\textwidth}
        \centerline{\includegraphics[scale=0.3]{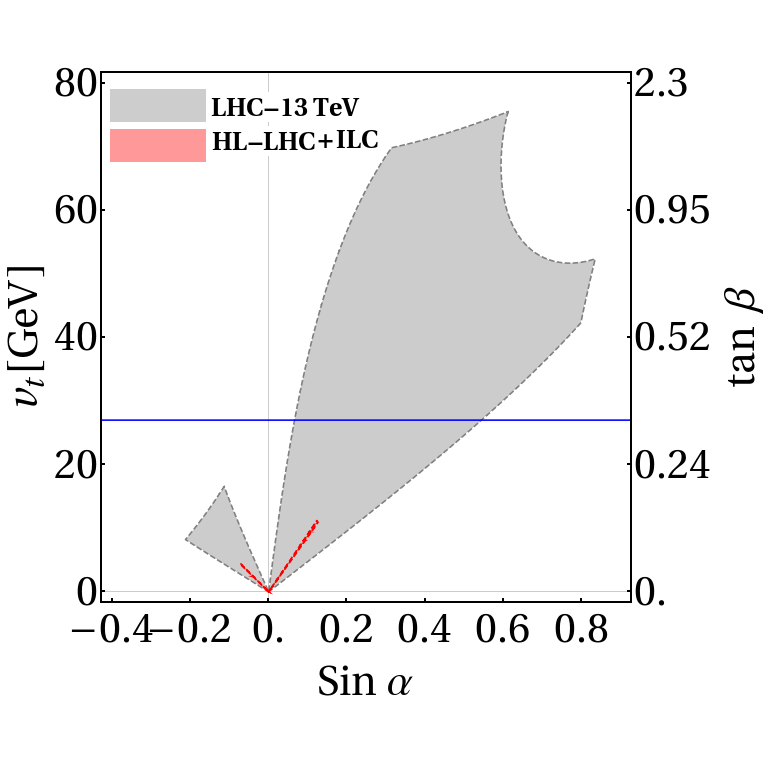}}
        \caption{\em Shrinking of the allowed region in Fig.~\ref{f:kvkf} when combined
            with the constraints from $\kappa_\gamma$ in the limit $M_1,M_2=0$. The shaded
            regions in gray and red characterize the bounds arising from current\cite{CMS:2018lkl}
            and future\cite{Fujii:2015jha} measurements of $\kappa_\gamma$ respectively. In
            drawing this plot, a mass hierarchy $m_3\approx m_5 \gg m_h$ has been
            assumed.
            The horizontal line corresponds to the lower limit on $v_t$ arising from
            the theoretical constraints.}
        \label{f:kvkfkg}
    \end{minipage}
\end{figure}

Let us now proceed by defining the Higgs coupling modifiers as:
\begin{eqnarray}
\kappa_x &=& \frac{g_{hxx}}{(g_{hxx})_{\rm SM}} \,,
\label{e:kappax}
\end{eqnarray}
where $x$ stands for the gauge bosons and the massive fermions.
In the GM models, these coupling modifiers can be calculated to be
\begin{subequations}
\begin{eqnarray}
\kappa_f &=& \frac{\cos \alpha}{\cos \beta} \,, \qquad (f={\rm generic ~ massive
~ fermion}) \,,
\label{e:kf} \\
\kappa_V &=& \cos \alpha \cos \beta + \frac{2 \sqrt{2}}{\sqrt{3}} 
\sin \alpha \sin \beta \,, \qquad (V=W,Z) \,. \label{e:kv} 
\end{eqnarray}
\label{e:kfkV}
\end{subequations}
Using the current limits on $\kappa_f$ and $\kappa_V$\cite{CMS:2018lkl}, 
we display the $2\sigma$
allowed region (light gray shade) in the $\sin\alpha$-$v_t$ plane in Fig.~\ref{f:kvkf}.
In the same plot we also show how this allowed region will shrink if
the Higgs data continue to drift towards the SM expectations with the projected
accuracy at the HL-LHC (2\%) (red shade) and at the ILC ($< 1\%$) (green shade)~\cite{Fujii:2015jha}. 
Clearly, with the projected accuracies at the
HL-LHC and the ILC, we can hope to constrain $v_t$ to be smaller than $20$~GeV.

A crucial observation of our paper is that the allowed region in Fig.~\ref{f:kvkf} from
$\kappa_f$ and $\kappa_V$ can be considerably reduced further by superimposing the
constraints from $h\to\gamma\gamma$ even in the limit when the charged
scalars are super heavy. To illustrate, let us first write down the effective
$h\gamma\gamma$ coupling as follows:
\begin{eqnarray}
{\mathscr L}_{h\gamma\gamma} =
g_{h\gamma\gamma} F_{\mu\nu} F^{\mu\nu} h \,,
\end{eqnarray}
where $F_{\mu\nu}= \partial_\mu A_\nu-\partial_\nu A_\mu$ is the usual
electromagnetic field tensor. Now we define the $h\gamma\gamma$ coupling
modifier as
\begin{eqnarray}
\kappa_\gamma = \frac{g_{h\gamma\gamma}}{(g_{h\gamma\gamma})_{\rm SM}} \,,
\end{eqnarray}
whose analytic expression for the GM model can be written as\footnote{
The factor $5$ in front of the scalar loop contribution ${\cal A_S}(\tau_5)$ arises from the sum of the  singly and doubly charged scalar contributions from the custodial fiveplet. }
\begin{eqnarray}
\kappa_\gamma = \frac{\kappa_V {\cal A_W}(\tau_W) + \sum_{f} Q_f^2 N^f_c \kappa_f {\cal A}_f (\tau_f) + \kappa_3 {\cal A_S}(\tau_3) + 5 \kappa_5 {\cal A_S}(\tau_5) }{{\cal A_W}(\tau_W) + \sum_{f} Q_f^2 N^f_c {\cal A}_f (\tau_f) } \,,
\label{e:kapgam}
\end{eqnarray}
where, $Q_f$ and $N^f_c$ denote the electric charge and the color factor respectively for the fermion, $f$, and,
  defining $\tau_x = (2m_x/m_h)^2$, the loop functions are given by \cite{Gunion:1989we},
\begin{subequations}
	\begin{eqnarray}
	{\cal A_W}(\tau_W) &=& 2 + 3 \tau_W + 3\tau_W(2 - \tau_W)f(\tau_W) \,, \label{AW} \\
	{\cal A}_f(\tau_f) &=& -2\tau_f\left[1 + (1-\tau_f) f(\tau_f)\right] \,, \label{Af}\\
	{\cal A_S}(\tau_i) &=& -\tau_i \left[1 - \tau_if(\tau_i)\right] \quad (i=3,5) \,. \label{AS} 
	\end{eqnarray}
	\label{e:loopfactors}
\end{subequations}
In \Eqn{e:loopfactors} the function $f(\tau)$ is defined as
\begin{eqnarray}
f(\tau) &=& \begin{cases} \left[\sin^{-1}\left(\sqrt{\frac{1}{\tau}}\right)\right]^2 \,, \qquad \qquad \qquad
{\rm for} ~~  \tau > 1 \,, \\
-\frac{1}{4} \left[ \log\left[\frac{1 + \sqrt{1 - \tau}}{1 - \sqrt{1 - \tau}}\right] - i \pi \right]^2 \,, \qquad
{\rm for} ~~  \tau \leq 1 \,.
\end{cases}
\end{eqnarray}
The expressions for $\kappa_3$ and $\kappa_5$ appearing in \Eqn{e:kapgam},
which encapsulate the contributions from the charged scalars that are members
of the custodial triplet and fiveplet respectively, are given by\footnote{
Note that the equality $g_{hH_5^{+}H_5^{-}}=g_{hH_5^{++}H_5^{--}}$
follows from the custodial symmetry.}
\begin{figure}[htbp!]
	\centering
	\includegraphics[scale=0.3]{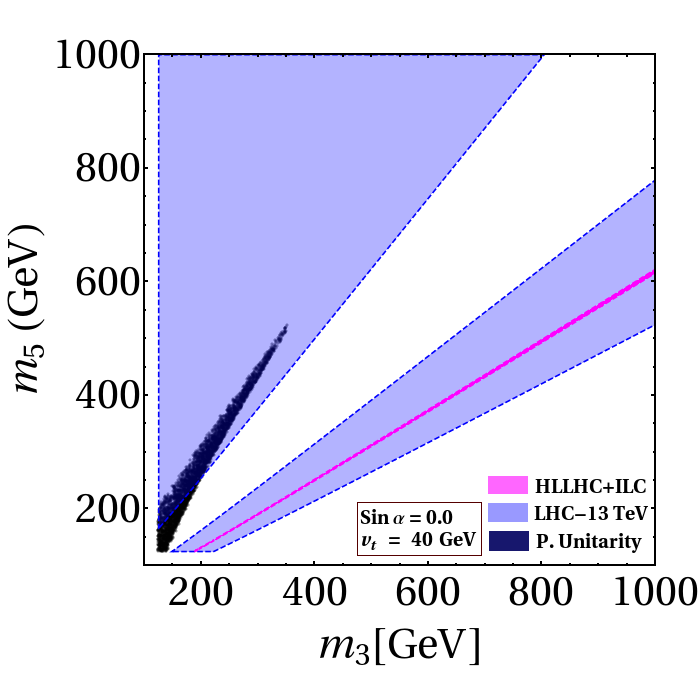} ~~~
	\includegraphics[scale=0.3]{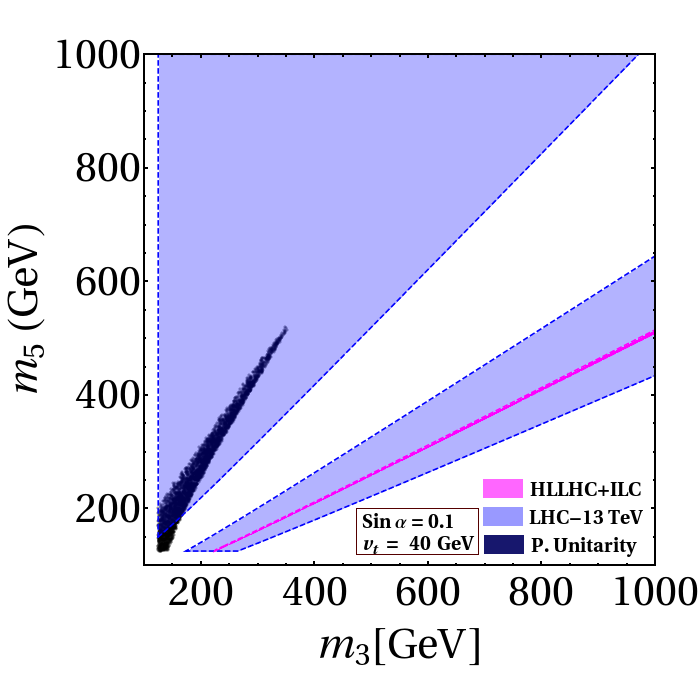} 
	\caption{\it Allowed regions at 95\% C.L. in the $m_3$-$m_5$ plane using the current
	(shaded in blue)\cite{CMS:2018lkl} and projected (shaded in magenta)\cite{Fujii:2015jha} accuracies in the measurement of $\kappa_\gamma$ for different sets of $\{\sin\alpha,~ v_t\}$
	values. The dark blue points cover the region that passes the theoretical
    constraints of perturbative unitarity.}
	\label{f:m3m5}
\end{figure}
\begin{subequations}
\label{e:k35}
\begin{eqnarray}
\kappa_3 &=&   \frac{M_W}{g\, m_3^2} g_{hH_3^+H_3^-} \,, \\
\kappa_5  &=&  \frac{M_W}{g\, m_5^2} g_{hH_5^{+}H_5^{-}} = \frac{M_W}{g\, m_5^2} g_{hH_5^{++}H_5^{--}} \,,
%
\end{eqnarray}
\end{subequations}
where, $g$ is the usual $SU(2)_L$ gauge coupling and $g_{h H_i^{+(+)} H_i^{-(-)}}$ denotes the trilinear 
coupling of $h$ with the $H_i^{+(+)} H_i^{-(-)}$ pair. The loop functions in \Eqn{e:loopfactors} saturate
to finite nonzero values once the particles circulating in the loop become much heavier than $m_h$.
Therefore, the (non)decoupling behaviors of the charged scalars will be captured by the quantities
$\kappa_3$ and $\kappa_5$. In the limit $M_1,~ M_2 =0$~\cite{Chanowitz:1985ug,Gunion:1989ci,Englert:2013zpa,Englert:2013wga}, we can calculate
\begin{subequations}
	\begin{eqnarray}
	\kappa_3 &=& \frac{-1}{2 m_3^2} \left[2 m_3^2 \left(\cos \alpha \cos \beta + \frac{2 \sqrt{2}}{\sqrt{3}} \sin \alpha \sin \beta \right)+2 m_h^2 \left( \frac{2 \sqrt{2}}{\sqrt{3}} \frac{\sin \alpha \cos^2 \beta}{\sin \beta} + \frac{\cos \alpha \sin^2\beta}{\cos \beta} \right)\right] \,, \label{e:k3} \\
	\kappa_5 &=&  \frac{-1}{2 m_5^2} \left[ \frac{2 \sqrt{2}}{\sqrt{3}} \frac{\sin \alpha }{ \sin \beta} \left(m_h^2+2 m_5^2\right)+ m_3^2 \left( 6 \cos \alpha \cos \beta - 4\sqrt{6} \frac{\sin \alpha \cos^2 \beta}{\sin \beta} \right) \right] \,. \label{k5}
	\end{eqnarray} 
	\label{e:kappach}
\end{subequations}
Evidently, in the limit $m_3\approx m_5 \gg m_h$, both $\kappa_3$ and $\kappa_5$ assume nonzero values,
which implies that contrary to the common perception\cite{Hartling:2014zca} the charged scalars do not
decouple. Note that for $M_1,M_2 =0$ but $v_t\ne 0$, the charged scalars receive their masses purely from
the SSB sources and consequently, such nondecoupling behaviors are to be expected\cite{Bhattacharyya:2014oka}.
Because of this, the allowed region in Fig.~\ref{f:kvkf} can be further constrained by superimposing
the bounds from $\kappa_\gamma$, which has been shown in Fig.~\ref{f:kvkfkg}. From Fig.~\ref{f:kvkfkg} we can easily
see that, for $M_1, M_2=0$, using the projected combined accuracy for the HL-LHC and the ILC~\cite{Fujii:2015jha}, we can restrict the 
upper limit on $v_t$ to be as low as $10$~GeV at $95\%$ C.L.
From Fig.~\ref{f:kvkfkg} it is also clear that the values of $\sin\alpha$ and $v_t$
can be strongly correlated if the measurement of $\kappa_\gamma$ agrees with
the SM with the expected combined precision at the HL-LHC and the ILC.
Consequently, the bound on $v_t$ will foster a strong upper limit on $\sin \alpha~ (|\sin \alpha|< 0.1) $.
Quite clearly, when combined with the lower limit on $v_t$ stemming from perturbative
unitarity, we can potentially rule out the variants of GM models with
$M_1,M_2=0$ using future Higgs precision data.
Finally, in Fig.~\ref{f:m3m5}, we display the allowed parameter space at $95\%$ C.L. in the $m_3$-$m_5$ plane
using the current as well as future precisions in $\kappa_\gamma$ for 
different sets of $\{ \sin \alpha, v_t \}$ values. We find that as the accuracy in $\kappa_\gamma$ improves,
the allowed region shrinks rapidly to the extent that with the future precision of about $1\%$
 at the HL-LHC and the ILC\cite{Fujii:2015jha}, the allowed area
reduces to a thin sliver in the parameter space, thereby allowing us to strongly
correlate $m_3$ and $m_5$ as well.
In the same figure we also show how perturbative unitarity entails a
{\em different} correlation in the $m_3$-$m_5$ plane (dark blue points)
in the limit $M_1,M_2=0$. This implies that if the measurement of
$h\to \gamma\gamma$ signal strength continues to agree with the
corresponding SM expectation with higher precision, the GM models
with $M_1,M_2=0$ will be increasingly disfavored.
 We have checked that the allowed region 
from $\kappa_\gamma$ in Fig.~\ref{f:m3m5}
does not crucially depend on the value of $v_t$ as long as it is nonzero.

To summarize, we have revisited the scalar sector of the GM model in the light of Higgs
precision data.
The GM model has long been preferred over minimal triplet model due to
its attribution towards the preservation of custodial symmetry at the tree-level
and thus lifting the stringent bound on the triplet scalar VEV. In this paper,
we have demonstrated that, for certain variants of the GM models, the triplet VEV can still
be tightly constrained using the Higgs precision data, especially the measurement of
the $h\to \gamma\gamma$ signal strength. More specifically, we have shown that when
the trilinear couplings in the scalar potential of \Eqn{e:potential} vanish, the
charged scalars do not decouple from the $h\to \gamma\gamma$ decay amplitude.
Consequently, using the projected precision in the measurement of $\kappa_\gamma$,
we have been able to set an upper limit on the triplet VEV, $v_t$, which can be
as low as $10$~GeV.
This nondecoupling behavior of the charged scalars in the context of the GM model
does not appear to be a widespread knowledge and therefore constitutes a new
finding of our paper.
 The upshot of our analysis is that, unlike the bound on $v_t$
obtained using the flavor data\cite{Hartling:2014aga,Biswas:2018jun}, our bounds derived from $\kappa_\gamma$
do not get washed away when the charged scalars are super heavy. We have also
demonstrated how, using a precise measurement of $\kappa_\gamma$, the masses of
the custodial triplet and fiveplet scalars can be correlated.
Moreover, when these bounds are used in conjunction with the theoretical constraints
from perturbative unitarity and stability, we have shown how the variants
of GM models with $M_1,M_2=0$ can fall out of favor in the near future.
Thus our current study highlights the importance of the measurement of Higgs to
diphoton signal strength in the future experiments in the context of GM models.

\section*{Acknowledgements}
The work of DD has been supported by the Swedish Research Council,
contract number 2016-05996.


\bibliographystyle{JHEP}
\bibliography{references.bib}

\providecommand{\href}[2]{#2}\begingroup\raggedright\begin{thebibliography}{10}

\bibitem{Aad:2012tfa}
{\scshape ATLAS} collaboration, G.~Aad et~al., \emph{{Observation of a new
  particle in the search for the Standard Model Higgs boson with the ATLAS
  detector at the LHC}},
  \href{http://dx.doi.org/10.1016/j.physletb.2012.08.020}{\emph{Phys. Lett.}
  {\bf B716} (2012) 1--29}, [\href{http://arxiv.org/abs/1207.7214}{{\tt
  1207.7214}}].

\bibitem{Chatrchyan:2012xdj}
{\scshape CMS} collaboration, S.~Chatrchyan et~al., \emph{{Observation of a new
  boson at a mass of 125 GeV with the CMS experiment at the LHC}},
  \href{http://dx.doi.org/10.1016/j.physletb.2012.08.021}{\emph{Phys. Lett.}
  {\bf B716} (2012) 30--61}, [\href{http://arxiv.org/abs/1207.7235}{{\tt
  1207.7235}}].

\bibitem{Georgi:1985nv}
H.~Georgi and M.~Machacek, \emph{{DOUBLY CHARGED HIGGS BOSONS}},
  \href{http://dx.doi.org/10.1016/0550-3213(85)90325-6}{\emph{Nucl. Phys.} {\bf
  B262} (1985) 463--477}.

\bibitem{Chanowitz:1985ug}
M.~S. Chanowitz and M.~Golden, \emph{{Higgs Boson Triplets With M ($W$) = M
  ($Z$) $\cos \theta \omega$}},
  \href{http://dx.doi.org/10.1016/0370-2693(85)90700-2}{\emph{Phys. Lett.} {\bf
  165B} (1985) 105--108}.

\bibitem{Gunion:1989ci}
J.~F. Gunion, R.~Vega and J.~Wudka, \emph{{Higgs triplets in the standard
  model}}, \href{http://dx.doi.org/10.1103/PhysRevD.42.1673}{\emph{Phys. Rev.}
  {\bf D42} (1990) 1673--1691}.

\bibitem{Kanemura:2012rs}
S.~Kanemura and K.~Yagyu, \emph{{Radiative corrections to electroweak
  parameters in the Higgs triplet model and implication with the recent Higgs
  boson searches}},
  \href{http://dx.doi.org/10.1103/PhysRevD.85.115009}{\emph{Phys. Rev.} {\bf
  D85} (2012) 115009}, [\href{http://arxiv.org/abs/1201.6287}{{\tt
  1201.6287}}].

\bibitem{Hartling:2014aga}
K.~Hartling, K.~Kumar and H.~E. Logan, \emph{{Indirect constraints on the
  Georgi-Machacek model and implications for Higgs boson couplings}},
  \href{http://dx.doi.org/10.1103/PhysRevD.91.015013}{\emph{Phys. Rev.} {\bf
  D91} (2015) 015013}, [\href{http://arxiv.org/abs/1410.5538}{{\tt
  1410.5538}}].

\bibitem{Biswas:2018jun}
A.~Biswas, D.~K. Ghosh, A.~Shaw and S.~K. Patra, \emph{{$b \to c \ell \nu$
  anomalies in light of extended scalar sectors}},
  \href{http://arxiv.org/abs/1801.03375}{{\tt 1801.03375}}.

\bibitem{Logan:2010en}
H.~E. Logan and M.-A. Roy, \emph{{Higgs couplings in a model with triplets}},
  \href{http://dx.doi.org/10.1103/PhysRevD.82.115011}{\emph{Phys. Rev.} {\bf
  D82} (2010) 115011}, [\href{http://arxiv.org/abs/1008.4869}{{\tt
  1008.4869}}].

\bibitem{Kanemura:2013mc}
S.~Kanemura, M.~Kikuchi and K.~Yagyu, \emph{{Probing exotic Higgs sectors from
  the precise measurement of Higgs boson couplings}},
  \href{http://dx.doi.org/10.1103/PhysRevD.88.015020}{\emph{Phys. Rev.} {\bf
  D88} (2013) 015020}, [\href{http://arxiv.org/abs/1301.7303}{{\tt
  1301.7303}}].

\bibitem{Belanger:2013xza}
G.~Belanger, B.~Dumont, U.~Ellwanger, J.~F. Gunion and S.~Kraml, \emph{{Global
  fit to Higgs signal strengths and couplings and implications for extended
  Higgs sectors}},
  \href{http://dx.doi.org/10.1103/PhysRevD.88.075008}{\emph{Phys. Rev.} {\bf
  D88} (2013) 075008}, [\href{http://arxiv.org/abs/1306.2941}{{\tt
  1306.2941}}].

\bibitem{Hartling:2014zca}
K.~Hartling, K.~Kumar and H.~E. Logan, \emph{{The decoupling limit in the
  Georgi-Machacek model}},
  \href{http://dx.doi.org/10.1103/PhysRevD.90.015007}{\emph{Phys. Rev.} {\bf
  D90} (2014) 015007}, [\href{http://arxiv.org/abs/1404.2640}{{\tt
  1404.2640}}].

\bibitem{Chiang:2014bia}
C.-W. Chiang, S.~Kanemura and K.~Yagyu, \emph{{Novel constraint on the
  parameter space of the Georgi-Machacek model with current LHC data}},
  \href{http://dx.doi.org/10.1103/PhysRevD.90.115025}{\emph{Phys. Rev.} {\bf
  D90} (2014) 115025}, [\href{http://arxiv.org/abs/1407.5053}{{\tt
  1407.5053}}].

\bibitem{Kanemura:2014bqa}
S.~Kanemura, K.~Tsumura, K.~Yagyu and H.~Yokoya, \emph{{Fingerprinting
  nonminimal Higgs sectors}},
  \href{http://dx.doi.org/10.1103/PhysRevD.90.075001}{\emph{Phys. Rev.} {\bf
  D90} (2014) 075001}, [\href{http://arxiv.org/abs/1406.3294}{{\tt
  1406.3294}}].

\bibitem{Chiang:2015kka}
C.-W. Chiang and K.~Tsumura, \emph{{Properties and searches of the exotic
  neutral Higgs bosons in the Georgi-Machacek model}},
  \href{http://dx.doi.org/10.1007/JHEP04(2015)113}{\emph{JHEP} {\bf 04} (2015)
  113}, [\href{http://arxiv.org/abs/1501.04257}{{\tt 1501.04257}}].

\bibitem{Logan:2015xpa}
H.~E. Logan and V.~Rentala, \emph{{All the generalized Georgi-Machacek
  models}}, \href{http://dx.doi.org/10.1103/PhysRevD.92.075011}{\emph{Phys.
  Rev.} {\bf D92} (2015) 075011}, [\href{http://arxiv.org/abs/1502.01275}{{\tt
  1502.01275}}].

\bibitem{Chiang:2015rva}
C.-W. Chiang, S.~Kanemura and K.~Yagyu, \emph{{Phenomenology of the
  Georgi-Machacek model at future electron-positron colliders}},
  \href{http://dx.doi.org/10.1103/PhysRevD.93.055002}{\emph{Phys. Rev.} {\bf
  D93} (2016) 055002}, [\href{http://arxiv.org/abs/1510.06297}{{\tt
  1510.06297}}].

\bibitem{Chiang:2015amq}
C.-W. Chiang, A.-L. Kuo and T.~Yamada, \emph{{Searches of exotic Higgs bosons
  in general mass spectra of the Georgi-Machacek model at the LHC}},
  \href{http://dx.doi.org/10.1007/JHEP01(2016)120}{\emph{JHEP} {\bf 01} (2016)
  120}, [\href{http://arxiv.org/abs/1511.00865}{{\tt 1511.00865}}].

\bibitem{Chang:2017niy}
J.~Chang, C.-R. Chen and C.-W. Chiang, \emph{{Higgs boson pair productions in
  the Georgi-Machacek model at the LHC}},
  \href{http://dx.doi.org/10.1007/JHEP03(2017)137}{\emph{JHEP} {\bf 03} (2017)
  137}, [\href{http://arxiv.org/abs/1701.06291}{{\tt 1701.06291}}].

\bibitem{Li:2017daq}
B.~Li, Z.-L. Han and Y.~Liao, \emph{{Higgs production at future
  e$^{+}$e$^{−}$ colliders in the Georgi-Machacek model}},
  \href{http://dx.doi.org/10.1007/JHEP02(2018)007}{\emph{JHEP} {\bf 02} (2018)
  007}, [\href{http://arxiv.org/abs/1710.00184}{{\tt 1710.00184}}].

\bibitem{Aoki:2007ah}
M.~Aoki and S.~Kanemura, \emph{{Unitarity bounds in the Higgs model including
  triplet fields with custodial symmetry}},
  \href{http://dx.doi.org/10.1103/PhysRevD.89.059902,
  10.1103/PhysRevD.77.095009}{\emph{Phys. Rev.} {\bf D77} (2008) 095009},
  [\href{http://arxiv.org/abs/0712.4053}{{\tt 0712.4053}}].

\bibitem{Chiang:2012cn}
C.-W. Chiang and K.~Yagyu, \emph{{Testing the custodial symmetry in the Higgs
  sector of the Georgi-Machacek model}},
  \href{http://dx.doi.org/10.1007/JHEP01(2013)026}{\emph{JHEP} {\bf 01} (2013)
  026}, [\href{http://arxiv.org/abs/1211.2658}{{\tt 1211.2658}}].

\bibitem{Degrande:2017naf}
C.~Degrande, K.~Hartling and H.~E. Logan, \emph{{Scalar decays to
  $\gamma\gamma$, $Z\gamma$, and $W\gamma$ in the Georgi-Machacek model}},
  \href{http://dx.doi.org/10.1103/PhysRevD.96.075013}{\emph{Phys. Rev.} {\bf
  D96} (2017) 075013}, [\href{http://arxiv.org/abs/1708.08753}{{\tt
  1708.08753}}].

\bibitem{CMS:2018lkl}
C.~Collaboration, \emph{{Combined measurements of the Higgs boson's couplings
  at $\sqrt{s}=13$ TeV}},  Tech. Rep. CMS-PAS-HIG-17-031, CMS, 2018.

\bibitem{Fujii:2015jha}
K.~Fujii et~al., \emph{{Physics Case for the International Linear Collider}},
  \href{http://arxiv.org/abs/1506.05992}{{\tt 1506.05992}}.

\bibitem{Gunion:1989we}
J.~F. Gunion, H.~E. Haber, G.~L. Kane and S.~Dawson, \emph{{The Higgs Hunter's
  Guide}}, {\emph{Front. Phys.} {\bf 80} (2000) 1--404}.

\bibitem{Englert:2013zpa}
C.~Englert, E.~Re and M.~Spannowsky, \emph{{Triplet Higgs boson collider
  phenomenology after the LHC}},
  \href{http://dx.doi.org/10.1103/PhysRevD.87.095014}{\emph{Phys. Rev.} {\bf
  D87} (2013) 095014}, [\href{http://arxiv.org/abs/1302.6505}{{\tt
  1302.6505}}].

\bibitem{Englert:2013wga}
C.~Englert, E.~Re and M.~Spannowsky, \emph{{Pinning down Higgs triplets at the
  LHC}}, \href{http://dx.doi.org/10.1103/PhysRevD.88.035024}{\emph{Phys. Rev.}
  {\bf D88} (2013) 035024}, [\href{http://arxiv.org/abs/1306.6228}{{\tt
  1306.6228}}].

\bibitem{Bhattacharyya:2014oka}
G.~Bhattacharyya and D.~Das, \emph{{Nondecoupling of charged scalars in Higgs
  decay to two photons and symmetries of the scalar potential}},
  \href{http://dx.doi.org/10.1103/PhysRevD.91.015005}{\emph{Phys. Rev.} {\bf
  D91} (2015) 015005}, [\href{http://arxiv.org/abs/1408.6133}{{\tt
  1408.6133}}].

\end{thebibliography}\endgroup
\end{document}